\documentclass[prb,twocolumn,superscriptaddress,showpacs,floatfix]{revtex4}
\usepackage{epsfig,graphicx,subfigure}
\newcommand\be{\begin{equation}}
\newcommand\ee{\end{equation}}
                                                                             
\begin{document}

\title{Mean field theory for double-perovskites}
\author{L. Brey}
\affiliation{Instituto de Ciencia de Materiales de Madrid,
  CSIC. Cantoblanco, 28049 Madrid, Spain}
\author{M.J. Calder\'on}
\affiliation{Condensed Matter Theory Center, Department of Physics,
  University of Maryland, College Park, Maryland 20742-4111}
\author{S. \surname{Das Sarma}}
\affiliation{Condensed Matter Theory Center, Department of Physics,
  University of Maryland, College Park, Maryland 20742-4111}
\author{F. Guinea}
\affiliation{Instituto de Ciencia de Materiales de Madrid,
  CSIC. Cantoblanco, 28049 Madrid, Spain}
\date{\today}

\begin{abstract}
A mean field approximation of a model for double
perovskites that takes into account the coupling between itinerant
electron spins and localized spins is developed. As in previously reported theoretical results,
and contrary to experimental observation,
the critical temperature is suppressed for large electron
density. 
An effective Heisenberg model reveals the cause of this
discrepancy: the competition between degenerate antiferromagnetic and
ferromagnetic channels. This degeneracy can be broken by the inclusion
of a Hubbard-type U term. 
It is therefore suggested that electron correlation
effects need to be
incorporated in the minimal model of double perovskites in order to
explain the experimental observation of increasing ferromagnetic
critical temperature with increasing electron doping.
\end{abstract}

\pacs{75.10.-b, 
75.47.Gk 
}

\maketitle

\section{Introduction}
\label{sec:introduction}
A large magnetoresistance is a desirable property for certain
technological applications such as writing and reading magnetic
memories. Manganites, the most studied colossal magnetoresistance
systems,~\cite{tokura00} have the drawback of having low ferromagnetic
transition (Curie)
temperatures $T_{\rm C}$ (compared to room temperature). They are half-metals,
namely, metallic for one spin orientation and insulating for the
other, but this property gets rapidly suppressed when increasing
temperature. Half-metalicity produces the low field (extrinsic) 
magnetoresistance
measured in polycrystalline manganites and the tunneling
magnetoresistance in artificially created barriers. 
Increasing the operation temperature
of these devices is a major issue that has lead to the search of other
half-metals with higher $T_{\rm C}$. One such example are
double perovskites of general formula A$_2$BB'O$_6$ (A=Sr,Ca,Ba,La,K, B=Fe,
B'=Mo,Re).~\cite{kobayashi98,kobayashi99,sarma01,saitoh02} 
Polycrystalline double perovskites show large
magnetoresistance  at low fields due to half-metalicity even at room temperature, 
 and their
$T_{\rm C}$ is above $400$K.~\cite{kobayashi98} Band structure
calculations~\cite{kobayashi98,sarma00,fang01,saitoh02,szotek03} reveal a $\sim 0.8$ eV
gap in the majority up spin bands at the Fermi level while the down
spin bands cross it.~\cite{sarma00}

\begin{figure}
        \includegraphics[clip,width=3in]{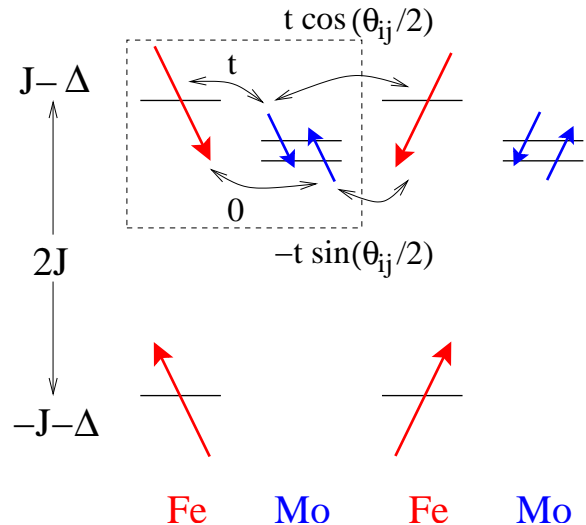}
        \caption{(color online) Description of the parameters of the model and
        definition of the paramagnetic state used. $t$ stands for
        $t_{\rm Fe-Mo}$. The level with energy $-J-\Delta$ correspond to the
        localized electrons in $Fe(3d^5)$ that are considered in the
        model as localized classical spins $S=5/2$. The dash box defines
        the unit cell $i$. 
        In the limit $J\rightarrow \infty$
        the spin of the conduction electron on the upper band is
        always strictly antiparallel to the localized spin.}
\label{fig:modelDE}
\end{figure}

The ordered double perovskite structure consists
of alternating BO$_6$ and B'O$_6$ octahedra in a cubic lattice. 
Anti-site disorder in
this lattice has strong effects in magnetic properties.~\cite{ray01,balcells01,navarro01,sanchez02,navarro03,sher05} 
Fe is in the trivalent state ($3d^5$)
with its five electrons
localized in the spin-up d-orbitals.~\cite{sarma00} Mo$^{5+}$ provides one
electron (4$d^1$) per Fe to the conduction band, electron density $c=1$, and Re$^{5+}$ provides
two (5$d^2$), $c=2$. These
electrons are moving in the Fe-Mo/Re hybrid band formed by
t$_{2g}$ spin-down orbitals.
The cubic symmetry causes a
splitting between the t$_{2g}$ and e$_g$ orbitals, such that t$_{2g}$
orbitals are lower in energy and the only ones occupied by the
conduction electrons.
The spin-down orbitals are well above the
spin-up orbitals due to the strong Hund's coupling in 
Fe (see Fig.~\ref{fig:modelDE}).
Fe is the magnetically active ion;
its five localized electrons render a local spin $S=5/2$. On the other
hand, Mo and Re are paramagnetic ions.~\cite{ray01} The
theoretical magnetization per formula unit is 4$\mu_B$ though smaller
values have been measured probably due to anti-site disorder, as
nearest neighbor Fe-Fe
superexchange (SE) interaction is antiferromagnetic.~\cite{sanchez02,navarro03,balcells01} In an ordered lattice, Fe-Fe SE is
very weak as Fe ions are too far away ($>5.5 \AA$) for this
interaction to be important. 
Doping on the A site changes the density of conduction electrons
per Fe: substituting a trivalent ion (e.g. La) for Sr
(Sr$_{2-x}$La$_x$FeMoO$_6$) increases the electron density to $c=1+x$ while a
monovalent ion (e.g. K) gives $c=1-x$.

Superexchange interaction, which is extremely weak in double
perovskites, has been ruled out as the cause of magnetic
ordering in these materials. Instead, there is experimental
evidence for the existence of two sublattices,~\cite{tovar02} the
localized spins in Fe and the delocalized electrons, that interact
antiferromagnetically due to the strong Hund's coupling $J$ on Fe
ions. Other materials that present these two coupled sublattices are
manganites~\cite{tokura00} and diluted magnetic semiconductors such as
Ga$_{1-x}$Mn$_x$As.~\cite{dassarma03GaMnAs,dassarma03GaMnAsSSC} In
manganites the conduction electrons are ferromagnetically coupled to
the localized Mn spins that form a pseudocubic lattice. Hund's
coupling in manganites is very large and, in the limit
$J \rightarrow \infty$, leads to the double exchange (DE) mechanism
producing ferromagnetism.   
DE basically implies that the spin of the conduction
electron follows the orientation of the local spin. 
As hopping does not flip the spin, the kinetic
energy of the conduction electrons is minimized when the local spins
are all parallel to each other.~\cite{zener}  Hund's coupling in
double perovskites is antiferromagnetic,~\cite{tovar02}
 as the spin-up Fe levels are fully occupied,
 but it can equally lead to ferromagnetic order. However, unlike in
 manganites, the strong Hund's coupling only applies to every other
 site, since Mo is paramagnetic~\cite{ray01} and, therefore, the
 minimal model for double-perovskites is different from the simple DE
 in manganites.~\cite{sarma00}

Susceptibility measurements of
double-perovskites in the paramagnetic regime give a positive
Curie-Weiss parameter (i.e. ferromagnetic
interaction).~\cite{martinez00,navarro01,tovar02} 
This rules out a superexchange scenario, 
since SE would only give ferromagnetic Fe ordering
through antiferromagnetic coupling  between Fe:3d$^5$ and Mo:3d$^1$
({\it ferrimagnetic} ordering) and in
this case a negative Curie-Weiss parameter should be observed.
A conduction electron mediated ferromagnetism is suggested by the
observation that 
the strength of the 
ferromagnetic coupling and thus, the $T_{\rm C}$, can be increased by
electron doping.~\cite{navarro01PRB,navarro03,frontera03,fontcuberta05} In turn,
the increase
in $T_{\rm C}$ has been shown to be accompanied by an increase of the
density of states at the Fermi level.~\cite{navarro04}

A model for double perovskites that takes into account the coupling 
between itinerant
electrons and localized spins on Fe has previously been studied
theoretically by means of dynamical mean-field theory~\cite{millis03}
and Monte-Carlo simulations.~\cite{guinea03} Their common result is
that the critical temperature $T_{\rm C}$ is suppressed as the electron density
increases above a certain value. This is clearly in contradiction with experimental
observations of increasing $T_{\rm C}$ by electron doping. On the other
hand, similar models for perovskite manganites and diluted magnetic
semiconductors (GaMnAs) find that the $T_{\rm C}$ increases with the
density of states.


In this paper a mean field theory approximation is developed on this 
model for ordered double perovskites. The previously published
results~\cite{millis03,guinea03} are recovered in the appropriate
limits. 
As mentioned already,
these theoretical results are not consistent with experimental data. 
The advantage of the mean field theory developed here is that the
reason for the failure of this model becomes transparent. The Hamiltonian
can be written as an effective Heisenberg model with ferromagnetic and
antiferromagnetic terms. Due to the
degeneracy of the spin up and down levels in the paramagnetic atom,
these two channels compete resulting in the suppression
of $T_{\rm C}$ for large enough values of the electron density. 
We show how the intraband Coulomb repulsion $U$, which penalizes
the occupation of two spin orientations at the same site, can
prefer the FM channel over the AF leading to an increase of $T_{\rm C}$
with doping, as observed experimentally. This work will not take into account
the possibility that the Coulomb repulsion also induces orbital order.~\cite{TG04}

This paper is organized as follows. In Sec. II the minimal model for
double perovskites is
introduced. In Sec. III,  the mean field approximation is described,
the model is written as an effective Heisenberg model, and the  
effects of adding electron correlation effects
are analyzed. In Sec. IV, an alternative mean field calculation is
described. This approach integrates out the effect of the paramagnetic
sites and considers only the Fe sites, making a clear connection with
the double exchange model used in the description of manganites.
We conclude in Sec. V.


\section{Model}
\label{sec:model}

The full Hamiltonian can be written as the sum of two terms
\be
H=H_{\rm KE}+H_{\rm on-site}.
\ee
The first term is the kinetic energy of the conduction electrons.
The second term takes account of the strong Hund's coupling on Fe sites, $J$,
and
the difference in electronegativities, $\Delta$, between Fe and Mo/Re
sites.

The conduction electrons move between t$_{2g}$ orbitals. These
orbitals have planar symmetry which implies hybridization
only takes place between t$_{2g}$ orbitals of the same symmetry.
Therefore, the kinetic energy consists of three degenerate, two
dimensional, tight binding
systems $xy$, $yz$, and $zx$,
$H_{\rm KE}=H_{xy}^{\rm KE}+H_{yz}^{\rm KE}+H_{zx}^{\rm KE}$. 
The matrix element $t_{\rm Fe-Mo}$ connects
the d$_{ab}$ ($a,b=x,y,z$) orbitals of nearest neighbors in the $ab$
plane. $t_{\rm Mo-Mo}$ connects nearest neighbors in the Mo
sublattice. $t_{\rm Fe-Fe}$ is expected to be very small due to the more
localized nature of the $3d$ states, and is neglected here.

The large local spins on Fe sites ($S=5/2$) can be considered
classical and are characterized by an angle $\theta_i$. 
Due
to the large value for the Hund's coupling, the spin of the conduction electron on Fe
follows adiabatically the classical local spin configuration.

\section{Mean field Approximation. Dispersive M{\small o} bands.}
\label{sec:mean-field}
$T_{\rm C}$ is calculated from the expansion of the Free
Energy $F$ in powers of the magnetization $m=\langle
\cos \theta_i \rangle$, which is very small close to $T_{\rm C}$.
$T_{\rm C}$ is defined by the condition $\partial ^2 F/\partial m^2=0$. 
In order to write the Free Energy, we have to define the
paramagnetic regime correctly. Diagonalization of the one-dimensional problem
shows that the density of
states has a gap in the paramagnetic regime. This gap is
related to the fact that Mo is paramagnetic and there are two
conducting channels: one through each spin orientation.

In the paramagnetic regime, the spins on the Fe become randomly
oriented. Mo is paramagnetic, allowing two spin states. To recover
the gap in the DOS, the orientation
of these two states in the Mo has to be referenced to a neighboring Fe spin,
as shown in Fig.~\ref{fig:modelDE}.  The Fe and Mo related in this way
constitute our unit cell. In this way, the spin of the electrons in
the Mo is chosen to be parallel or
antiparallel to the direction of the Fe spin in the same cell.
Inside the unit cell, the Fe-Mo hopping is $1$ (for parallel spins) or
$0$ (for anti-parallel spins) and between
different cells it is determined by the angle formed by Fe ions in neighboring
cells $\theta_{ij}=\theta_i-\theta_j$: $\cos
(\theta_{ij}/2) $ (spin $\uparrow$ channel) and
$\sin  (\theta_{ij}/2)$ (spin $\downarrow$
channel). Therefore, the
system consists of two different channels that could not be
distinguished if the relative orientation of the spins in Mo and Fe
were not correctly defined.
In the virtual crystal approximation, each site
sees an average of all the sites in the lattice so the relevant
coefficients of the hopping terms are the thermal averages 
$\langle\cos (\theta_{ij}/2)\rangle \equiv
\langle\cos (\theta/2) \rangle $ and 
$\langle \sin (\theta_{ij}/2) \rangle \equiv \langle 
\sin (\theta/2)\rangle$. In the paramagnetic
regime both are equal to $2/3$.

For each of the equivalent planes, the Hamiltonian can then be written
\begin{widetext}
\begin{eqnarray}
H_{xy}&=&(J-\Delta)\sum_i d_i^+ d_i 
+t_{\rm Mo-Mo} \sum_{\langle i,j \rangle} \left[ \left\langle \cos
{{\theta}\over{2}} \right\rangle (c^+_{i,p} c_{j,p} +c^+_{i,ap}
c_{j,ap})
-\left\langle \sin
{{\theta}\over{2}} \right\rangle (c^+_{i,ap} c_{j,p} +c^+_{i,p}
c_{j,ap}) \right] \nonumber \\
&+&t_{\rm Fe-Mo} \sum_i d_i^+ \left[c_{i,p} 
+\left\langle \cos
{{\theta}\over{2}}  \right\rangle (c_{i-x,p}+c_{i-x-y,p}+c_{i-y,p})
-\left\langle \sin
{{\theta}\over{2}}  \right\rangle
(c_{i-x,ap}+c_{i-x-y,ap}+c_{i-y,ap})\right] \, ,
\label{eq:hxy}
\end{eqnarray}
\end{widetext}
where $c_{i,p(ap)}$ destroys an electron in Mo at cell $i$ with spin
parallel (antiparallel) to the spin of the Fe core spin, and $d_i$
destroys an electron in Fe at cell $i$, with the spin parallel to the
core spin. $H_{yz}$ and $H_{zx}$ have identical form and give the same
contribution to the total energy.

Close to the magnetic transition, we can write the hopping
coefficients $\langle\cos (\theta/2) \rangle$ and 
$\langle \sin (\theta/2) \rangle$ as an expansion in 
$m$ to second order
\begin{eqnarray}
\left\langle \cos {{\theta}\over{2}} \right\rangle &  \sim &  \frac{2}{3}+\frac{2}{5} m ^2
 \nonumber \\
\left\langle \sin  {{\theta}\over {2}}\right\rangle  & \sim &
\frac{2}{3}-\frac{2}{5} m^2 \, ,
\end{eqnarray}
where we are following the same procedure as in
Ref.~\onlinecite{degennes}.~\cite{DMFT} Using these expressions in the Hamiltonian in Eq.~(\ref{eq:hxy}), and
taking into account that there are three equivalent 2-dimensional bands,
the kinetic energy can be calculated as a function of $m$:
$E_{\rm KE}=E_{\rm KE}^0+\chi m^2$. Knowing that the entropy of the spin system
is
\begin{equation}
TS = \frac{1}{\beta} \left (  \log \left ( 2
\frac{\sinh ( \beta h )}{\beta h } \right )  +m \beta h  \right ),
\end{equation}
where $h$ is an external magnetic field and $\beta= k_{\rm B} T_{\rm C}$, and that $\partial ^2 F/\partial
m^2=0$, we get
$T_{\rm C}=2/3 \,[\partial  E_{\rm KE}/\partial (m^2)]$~[\onlinecite{degennes,GGA00}]
(see Appendix \ref{appendix1}).

\begin{figure}
        \includegraphics[clip,width=3in]{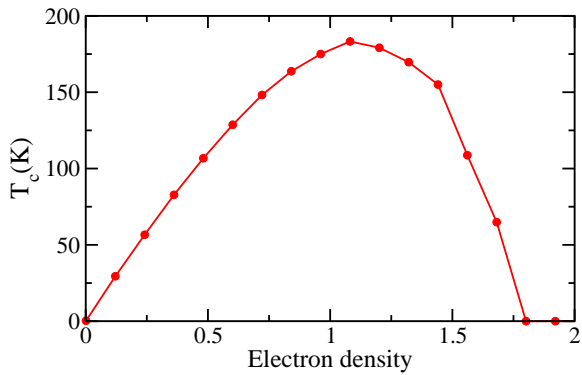}
        \caption{(color online) $T_{\rm C}$ versus electron
        density $c$ for a particular
        set of parameters $J-\Delta=0.3 eV$, $t_{\rm Mo-Mo}=0.15
eV$, and $t_{\rm Fe-Mo}=0.3 eV$. For all the range of parameters studied,
        there is a peak in $T_{\rm C}$ around $c \sim 1$ and $T_{\rm C}=0$ at $c
        \sim 2$. }
\label{fig:resultsU0} 
\end{figure}

Numerical results for $T_{\rm C}$ with parameters $J-\Delta=0.3$ eV,
$t_{\rm Mo-Mo}=0.15$
eV, and $t_{\rm Fe-Mo}=0.3$ eV are shown in
Fig.~\ref{fig:resultsU0}. These parameters are consistent with
ab-initio calculations~\cite{sarma00} and similar to the ones used in
previous theoretical works on this model.~\cite{millis03,guinea03}
$J+\Delta$ is considered to be infinite as the transitions to the
Fe spin level parallel to the localized spin involve very large energies.~\cite{millis03}
The results are in agreement with
Monte-Carlo~\cite{guinea03} and dynamical mean-field theory~\cite{millis03} calculations but in 
disagreement with experiments. In general,
the calculated $T_{\rm C}$ is lower
than that measured in experiments and shows a maximum around $c \sim 1$, while
it is suppressed for larger electron densities $c \sim 2$. This
behavior persists for a wide range of parameters ($0<t_{\rm Mo-Mo}< 0.25$ eV,
$0<t_{\rm Fe-Mo}< 0.5$ eV, and $-1.5$eV $< J-\Delta <1.5$eV).

The maximum of $T_{\rm C}$ around $c=1$ and its suppression around $c=2$ is
related to the form of the density of states. Around $c=1$ there is
a maximum on the DOS and, close to $c=2$ both parallel and
anti-parallel spin bands from Mo are filled. The effect of this
filling is more easily understood by introducing an effective Heisenberg model.

\subsection{Effective Heisenberg model}
To analyze the source of the discrepancy between theory and experiment 
the energy is written as a function of the relative angle between
neighboring spins 
\begin{eqnarray}
\Delta E&=&-\sum_{\langle i,j \rangle}\left( J_C^{\rm Fe-Mo}
\cos{{\theta_{ij}}\over{2}}+J_C^{\rm Mo-Mo}
\cos{{\theta_{ij}}\over{2}} \right. \nonumber \\ 
&+&\left. J_S^{\rm Fe-Mo}\sin{{\theta_{ij}}\over{2}}+
J_S^{\rm Mo-Mo}\sin{{\theta_{ij}}\over{2}} \right) \, ,
\label{eq:eff-heisenberg}
\end{eqnarray}
where the $J$'s are the expectation values of the operator pairs in
Eq.~(\ref{eq:hxy}).
In the $m \rightarrow 0$ limit, Eq.~(\ref{eq:eff-heisenberg}) is an effective Heisenberg model,

\begin{equation}
\Delta E^{\rm Heis}= -{\frac{1}{2 \sqrt{2}}}
 \sum_{\langle i,j \rangle}  
J_{\rm eff} \cos \theta_{ij} \, ,
\end{equation}
with
\be
J_{\rm eff}=J_C^{\rm Fe-Mo}+J_C^{\rm Mo-Mo}-J_S^{\rm Fe-Mo}-J_S^{\rm Mo-Mo}.
\label{eq:jeff}
\ee

\begin{figure}
 \includegraphics[clip,width=3.5in]{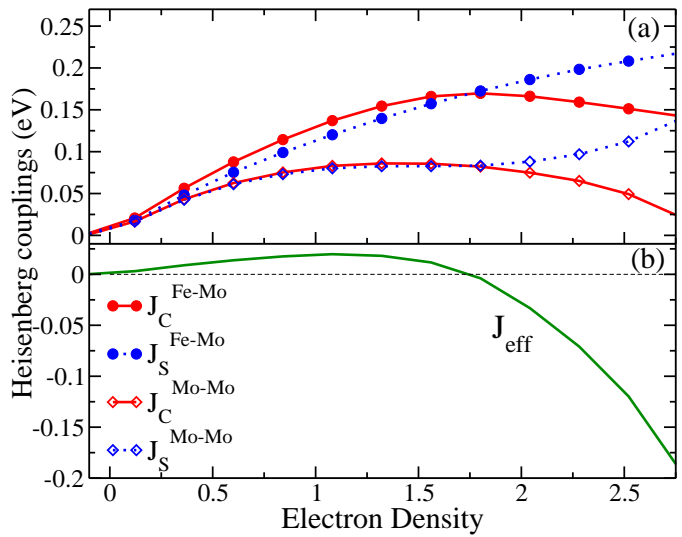}
        \caption{(color online) (a) Heisenberg coupling values for the ferromagnetic
        (solid) and
        antiferromagnetic (dashed) channels. The antiferromagnetic coupling
        gets stronger for large electron density $c>1.5$. (b) The total
        coupling $J_{\rm eff}$ as defined in Eq.~(\ref{eq:jeff}).}
\label{fig:heisenberg}
\end{figure}

Therefore, this effective Heisenberg model has competing ferromagnetic
($J_C$'s) and
antiferromagnetic ($J_S$'s) terms. In Fig.~\ref{fig:heisenberg}(a) the values of
the four different couplings are plotted as a function of electron density. 
As shown in
Fig.~\ref{fig:heisenberg}(b), the total coupling $J_{\rm eff}$ is
antiferromagnetic for large values of electron density. The electron
density at which $J_{\rm eff}$ becomes zero does not change significantly
within the range of 
the tight-binding parameters used. Therefore, we cannot expect to
override the suppression of $T_{\rm C}$ within this model.
The competition between the ferromagnetic and antiferromagnetic 
channels can lead to phase
separation,~\cite{guinea03} due to the fact that $J_{\rm eff}$ (and,
consequently, $T_{\rm C}$) depend strongly on the electron density.~\cite{GGA00}

 The ferromagnetic and antiferromagnetic channels are degenerate in energy since Mo is paramagnetic and both parallel and antiparallel states are
equally populated.

\subsection{Coulomb interaction}
It is clear from the previous section that the degeneracy of
ferromagnetic and antiferromagnetic channels needs to be broken to
obtain agreement with experimental results. Therefore, double
occupancy of a site must be penalized which is, of course, expected in
the presence of Coulomb interaction induced on-site electron
correlations (so far neglected in the theory). This can be done with an
intraband Hubbard term of the form

\be
H_U=U\sum_i n_{i\uparrow} n_{i\downarrow} \, .
\ee

\begin{figure}
        \includegraphics[clip,width=3in]{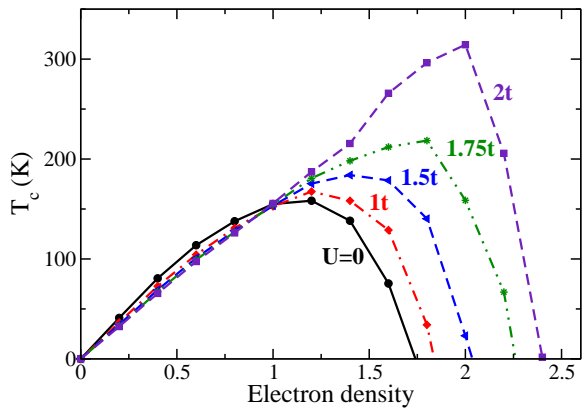}
        \caption{(color online) $T_{\rm C}$ versus electron density $c$ for different
          values of the Hubbard $U$. $J-\Delta=0.3$ eV, $t_{\rm Mo-Mo}=0.15$
 eV, and $t_{\rm Fe-Mo}=0.3$ eV. Moderate values of Coulomb
        repulsion $U$ produce a
        large increase in critical temperature and its suppression
        is shifted to higher electron density.}
\label{fig:resultsU}
\end{figure}

In Fig.~\ref{fig:resultsU}, $T_{\rm C}$ is plotted for different values
of $U$. Using relatively small values for $U$ ($U<<W\sim 8t$), we
observe that $T_{\rm C}$ is strongly enhanced and the position of
the maximum is shifted up from $c \sim 1$, consistent with experiment. 
\section{Mean Field Approximation. Non dispersive  M{\small o} bands.}
\label{sec:non-dispersive}
If we neglect the direct hopping between the Mo or Re orbitals, the connection between
the model studied here and the double exchange model used in the description
of the manganites becomes transparent. The Mo, Re orbitals can be replaced by
an energy dependent direct Fe-Fe hopping, and a correction to the energy of
the Fe orbitals. These quantities are:
\begin{eqnarray}
\epsilon_{\rm Fe} &= &
\epsilon^0_{\rm Fe} - \frac{t_{\rm Fe-Mo}^2}{\omega - (
  J - \Delta )} \nonumber \\
t_{\rm Fe-Fe} &= &
\frac{t_{\rm Fe-Mo}^2 \cos ( \theta_{ij} / 2 )}{\omega - (
  J - \Delta )} \,.
\label{tight_binding}
\end{eqnarray}
When the separation between the Fe and Mo levels, $J -
\Delta$, is large compared to the effective hybridization of the Fe
levels, namely, 
\begin{equation}
| J - \Delta | \ll \frac{t_{\rm Fe-Mo}^2}{| J - \Delta |} \, ,
\end{equation}
the model reduces to an effective double exchange model,~\cite{degennes} with hopping
integral:
\begin{equation}
t_{\rm eff} = \frac{t_{\rm Fe-Mo}^2 \cos (\theta_{ij} / 2)}{| J -
\Delta |} \,.
\end{equation}
A mean field solution to the problem can be obtained by making the
substitution:
\begin{equation}
\left\langle \cos \left( \frac{\theta_{ij}}{2} \right) \right\rangle
= \sqrt{\frac{1 + \langle m \rangle^2}{2}} \, ,
\label{DMFT_eq}
\end{equation}
where $m$ is the temperature dependent magnetization of the Fe ions. 

Keeping the full energy dependence of the tight binding parameters in
Eq.~(\ref{tight_binding}), one finds a self consistent solution for the band
structure, which can be written as a quadratic equation for the band
energies, $\epsilon^\pm_{k_x k_y}$. This equation can be solved, and we obtain:
\begin{widetext}
\begin{eqnarray}
\epsilon^\pm_{k_x k_y} &= &J + \frac{\Delta - J}{2} \pm \sqrt{
\frac{( \Delta - J )^2}{4} 
+ 4 t_{\rm Fe-Mo}^2 \left\{ 1 + \sqrt{\frac{1 + \langle m
\rangle^2}{2}} \left[ -1 + \left( \cos (
k_x ) + \cos ( k_y ) \right)^2 \right] \right\}} \, ,\nonumber \\
\epsilon^0_{k_x k_y} &= &\Delta \, .
\end{eqnarray}
\end{widetext}
These equations give the exact solution of the model at zero temperature
($m=1$), and they should describe qualitatively the changes in the electronic
structure induced by the magnetic fluctuations. Note that the absence of 
direct hopping between Mo orbitals leads to a dispersionless band which has
only weight at the Mo sites. 

From the knowledge of the dependence of the electronic bands as function of
the magnetization, one can calculate the electronic contribution to the free
energy, and obtain the value of the Curie temperature, as discussed in
Appendix A. We find:
\begin{widetext}
\begin{equation}
T_{\rm C} = \frac{2}{3} \frac{\partial^2 E_{\rm KE}}{\partial \langle m
\rangle^2}  =  \int \int_{\epsilon_{k} \le
\epsilon_F} d^2 k \sum_i   \frac{\pm t_{\rm Fe-Mo}^2 \left\{ \left[
\cos ( k_x ) + \cos ( k_y ) \right]^2 - 1 \right\}}{ \sqrt{
\frac{( \Delta - J)^2}{8} + 2 t_{\rm Fe-Mo}^2 \left\{ 1 + \sqrt{\frac{1}{2}}
\left[ -1 + \left( \cos ( k_x ) + \cos ( k_y ) \right)^2 \right] 
\right\}}} \, .\label{TC}
\end{equation}
\end{widetext}

The main drawback of this approximation is that it does not allow to
estimate the contribution of the direct hopping between Mo orbitals, $t_{\rm
  Mo-Mo}$, which is expected to be comparable to $t_{\rm Fe-Mo}$. These terms
tend to reduce the dependence of the electronic energy on the magnetization
of the Fe sites, lowering the value of $T_{\rm C}$. The qualitative
dependence of $T_{\rm C}$ on band filling obtained in this approximation is,
however, similar to the one considered in Sec.~\ref{sec:mean-field}, with $T_{\rm C} \rightarrow 0$ for electronic densities $c
\approx 2$. This fact can be explained, qualitatively, by noting that the
value of $T_{\rm C}$ in the effective double exchange model derived here
tends to have a maximum when the band is half filled. Because of the triple
degeneracy of the Mo orbitals, the corresponding density is $c =
1.5$.

\section{Conclusion}
\label{sec:conclusion}
We have developed a mean field theory for double perovskites,
e.g. Sr$_{2-x}$La$_x$FeMoO$_6$, within a minimal effective model
(Fig.~\ref{fig:modelDE}) including the strong Hund's coupling on the Fe sites and
the various contributions to the kinetic energy of electron hopping
between $t_{2g}$ orbitals through the Mo sites. Ferromagnetism arises
in the system due to constraints imposed on the hopping kinetic energy,
rather than due to Fe-Fe superexchange. Our simple mean field theory
reproduces the earlier theoretical results obtained with dynamical
mean field approximation~\cite{millis03} and direct Monte Carlo
simulations.~\cite{guinea03} Our theory gives a reasonable semi-quantitative
description of the observed Curie temperature in the double
perovskites for the `undoped' $x=0$ system where the carrier density
$c=1$ (per unit cell). But, for the doped double perovskites ($x \ne
0$) our mean field theory, along with the existing theories of
Refs.~\onlinecite{millis03} and \onlinecite{guinea03}, predicts a decreasing $T_{\rm C}$ with
increasing $x$ (with $c=1+x$ for electron doping and $c=1-x$ for hole
doping), which disagrees with experimental observations. The
experimental finding is that $T_{\rm C}$ increases with doping whereas the
theory finds a maximum around $c \approx 1$ (i.e. $x=0$).

We suggest, based in our mean-field formalism, that the experimental
observation of increasing $T_{\rm C}$ with doping may be an electron
correlation effect which opposes double occupancy of sites due to
intraband Coulomb repulsion. By introducing a simple Hubbard-U type
intraband correlation term, we qualitatively reproduce the
experimental trend of increasing Curie temperature with increasing
doping. In addition, the introduction of the Hubbard-U term also
enhances the $T_{\rm C}$ itself bringing theory and experiment into better
quantitative agreement. We therefore believe that strong correlation
effects are an inherent property of double perovskites. 

Finally, we discuss the possible connection between double perovskites
(DP) and diluted magnetic semiconductors (DMS),
e.g. Ga$_{1-x}$Mn$_x$As, from the perspective of magnetism. At first,
one notices some superficial similarities between DP and DMS
materials: both have optimal $T_{\rm C}$ of the order of a few hundreds of
Kelvin (although the highest reported $T_{\rm C}$ in GaMnAs is around
$200$K, substantially below the room temperature $T_{\rm C}$'s routinely
seen in DP materials); both manifest $T_{\rm C}$'s increasing with doping,
thereby indicating a role for carrier mediated ferromagnetism. There
are, however, important differences between DMS and DP magnetic
properties. In DMS, $T_{\rm C}=0$ for $x=0$ since Mn atoms serve the dual
roles of dopants (providing the carriers, which are holes for
Ga$_{1-x}$Mn$_x$As) and magnetic moments (i.e. the long range
ferromagnetic order arises from the order of the local Mn moments),
and therefore ferromagnetism vanishes in the absence of Mn. Thus the
$x=0$ situation in DMS is qualitatively similar to the $x=1$ DP
situation. The common
model~\cite{dassarma03GaMnAs,dassarma03GaMnAsSSC} adopted in the
literature to understand DMS ferromagnetism is a carrier-mediated
RKKY-Zener indirect exchange coupling between the Mn moments, with the
mean-field DMS $T_{\rm C}$ given by $T_{\rm C} \sim |J_{pd}|^2 x n^{1/3}$, where
$J_{pd}$ is the local `pd' exchange coupling between the Mn d-level
and the p-type holes (with density $n$) in the valence band of GaAs. 
This RKKY-Zener type DMS mean-field theory is obviously completely
inapplicable to the DP materials as it would predict an absurd DP
$T_{\rm C}$ of $10^5-10^6$K or larger (since both the magnetic moment
density and the carrier density are substantially higher in DP
materials than in DMS materials). We have developed the appropriate DP
mean-field theory in this paper with a reasonable $T_{\rm C} \sim
10^2-10^3$K.

It has, in fact, been suggested in the
literature~\cite{chattopadhyay01GaMnAs} that the DMS systems may actually be
closer to the non-perturbative double-exchange limit that the
perturbative RKKY limit. In such case, the DMS and the DP systems are
more similar in nature (and they are both then closer to manganites,
which are the quintessential double-exchange
materials~\cite{chattopadhyay01PRB,calderonMC98}). But, in this limit,
increasing doping should invariably lead to the eventual suppression
of $T_{\rm C}$, as we find in the theory developed in this paper.  
An important difference between DP and DMS
materials is, however, the fact that the DMS systems lose their
ferromagnetism (i.e. $T_{\rm C}$ becomes zero) for large values ($x \sim
0.1$) of Mn concentration. This also sharply distinguishes the DMS and
the DP materials. A natural question, based on our argument in favor
of the important role of a Hubbard type U-term in the DP materials is
whether such electron correlation effects are also important in DMS
materials. The answer to this question is not obvious at this
stage. One possibility is that correlation effects are completely
negligible in the DMS systems since the effective carrier density is
extremely low ($n \sim 10^{19}-10^{20}$ cm$^{-3}$ in DMS compared with
$10^{22}-10^{23}$ cm$^{-3}$ in DP materials), making the physics of
double occupancy irrelevant. Much more work will obviously be needed
to further understand the relationships and the differences in the
magnetic properties and mechanisms for various `oxide-type' magnetic
materials such as manganites, double perovskites, diluted magnetic
semiconductors, magnetically doped magnetic oxides
(e.g. Ti$_{1-x}$Co$_x$O$_2$), and even systems such as
Fe$_{1-x}$Co$_x$Si where correlation effects are thought to play an
important role.

\begin{acknowledgments}
This work is supported by MAT2005-07369-C03-03 (Spain) (LB) and
LPS-NSA, US-ONR, and NSF (MJC and SDS). 
\end{acknowledgments}

\appendix
\section{Mean field expression for the Curie temperature}
\label{appendix1}
The Free Energy of a system of classical spins of magnitude unity in
an external magnetic field $h$ is
\begin{equation}
F= - \frac{1}{\beta}\log \left ( 2 \frac{\sinh(\beta h )}{\beta h
} \right ) \,,
\end{equation}
from where the magnetization can be calculated as
\begin{equation}
m \equiv <m> = - \frac{\partial F }{\partial h } = \frac{1}{\tanh
( \beta h)} -\frac{1}{\beta h} \,. \label{magnetization}
\end{equation}

The entropy of the spin system is then
\begin{equation}
-TS= F - m h = \frac{1}{\beta} \left ( - \log \left ( 2
\frac{\sinh ( \beta h )}{\beta h } \right )  - m \beta h  \right ) \,.
\end{equation}
The total energy of the system, assuming the electrons are at zero
temperature and neglecting direct interactions between Fe spins,
can be written
\begin{equation}
F^{\rm total} = \chi m ^ 2 - T S = \chi m ^2 -\frac{1}{\beta} \left (
\log \left ( 2 \frac{\sinh ( \beta h )}{\beta h } \right )  + m
\beta h \right )\, ,
\end{equation}
where $\chi m^2$ is the kinetic energy of the conduction electrons. 

To obtain $h$, we minimize the total free energy with respect to
$h$, $\partial F ^{\rm total}/\partial h =0$. In the limit of
small $h$
\begin{eqnarray}
& & \frac{\sinh ( \beta h ) }{\beta h }  \equiv   1 + \frac{(
\beta h )
^2}{6} \nonumber \\
& & \log ( 2 \frac{\sinh ( \beta h ) }{\beta h })  \simeq \log (
2) +  \frac{( \beta h ) ^2 }{6}\,.
\end{eqnarray}
In this limit, the minimization condition gives
$\beta h = -3 m$,
and the free energy gets the form
\begin{equation}
F^{\rm total} = - \frac{1}{\beta} \log (2) + m ^2 \left ( \chi +
\frac{3}{2}k_{\rm B} T \right ).
\end{equation}
Therefore, the Curie temperature is~\cite{GGA00}
\begin{equation}
k_{\rm B} T_{\rm C} = - \frac{2}{3} \chi
\end{equation}

\bibliography{double-per}

\begin{thebibliography}{32}
\expandafter\ifx\csname natexlab\endcsname\relax\def\natexlab#1{#1}\fi
\expandafter\ifx\csname bibnamefont\endcsname\relax
  \def\bibnamefont#1{#1}\fi
\expandafter\ifx\csname bibfnamefont\endcsname\relax
  \def\bibfnamefont#1{#1}\fi
\expandafter\ifx\csname citenamefont\endcsname\relax
  \def\citenamefont#1{#1}\fi
\expandafter\ifx\csname url\endcsname\relax
  \def\url#1{\texttt{#1}}\fi
\expandafter\ifx\csname urlprefix\endcsname\relax\def\urlprefix{URL }\fi
\providecommand{\bibinfo}[2]{#2}
\providecommand{\eprint}[2][]{\url{#2}}

\bibitem[{\citenamefont{Tokura}(2000)}]{tokura00}
\bibinfo{editor}{\bibfnamefont{Y.}~\bibnamefont{Tokura}}, ed.,
  \emph{\bibinfo{title}{Colossal Magnetoresistance Oxides}}
  (\bibinfo{publisher}{Gordon and Breach}, \bibinfo{address}{New York},
  \bibinfo{year}{2000}).

\bibitem[{\citenamefont{Kobayashi et~al.}(1998)\citenamefont{Kobayashi, Kimura,
  Sawada, Terakura, and Tokura}}]{kobayashi98}
\bibinfo{author}{\bibfnamefont{K.-I.} \bibnamefont{Kobayashi}},
  \bibinfo{author}{\bibfnamefont{T.}~\bibnamefont{Kimura}},
  \bibinfo{author}{\bibfnamefont{H.}~\bibnamefont{Sawada}},
  \bibinfo{author}{\bibfnamefont{K.}~\bibnamefont{Terakura}}, \bibnamefont{and}
  \bibinfo{author}{\bibfnamefont{Y.}~\bibnamefont{Tokura}},
  \bibinfo{journal}{Nature} \textbf{\bibinfo{volume}{395}},
  \bibinfo{pages}{677} (\bibinfo{year}{1998}).

\bibitem[{\citenamefont{Kobayashi et~al.}(1999)\citenamefont{Kobayashi, Kimura,
  Tomioka, Sawada, Terakura, and Tokura}}]{kobayashi99}
\bibinfo{author}{\bibfnamefont{K.~I.} \bibnamefont{Kobayashi}},
  \bibinfo{author}{\bibfnamefont{T.}~\bibnamefont{Kimura}},
  \bibinfo{author}{\bibfnamefont{Y.}~\bibnamefont{Tomioka}},
  \bibinfo{author}{\bibfnamefont{H.}~\bibnamefont{Sawada}},
  \bibinfo{author}{\bibfnamefont{K.}~\bibnamefont{Terakura}}, \bibnamefont{and}
  \bibinfo{author}{\bibfnamefont{Y.}~\bibnamefont{Tokura}},
  \bibinfo{journal}{Phys. Rev. B} \textbf{\bibinfo{volume}{59}},
  \bibinfo{pages}{11159} (\bibinfo{year}{1999}).

\bibitem[{\citenamefont{Sarma}(2001)}]{sarma01}
\bibinfo{author}{\bibfnamefont{D.~D.} \bibnamefont{Sarma}},
  \bibinfo{journal}{Current Opinion in Solid State and Materials Science}
  \textbf{\bibinfo{volume}{5}}, \bibinfo{pages}{261} (\bibinfo{year}{2001}).

\bibitem[{\citenamefont{Saitoh et~al.}(2002)\citenamefont{Saitoh, Nakatake,
  Kakizaki, Nakajima, Morimoto, Xu, Moritomo, Hamada, and Aiura}}]{saitoh02}
\bibinfo{author}{\bibfnamefont{T.}~\bibnamefont{Saitoh}},
  \bibinfo{author}{\bibfnamefont{M.}~\bibnamefont{Nakatake}},
  \bibinfo{author}{\bibfnamefont{A.}~\bibnamefont{Kakizaki}},
  \bibinfo{author}{\bibfnamefont{H.}~\bibnamefont{Nakajima}},
  \bibinfo{author}{\bibfnamefont{O.}~\bibnamefont{Morimoto}},
  \bibinfo{author}{\bibfnamefont{S.}~\bibnamefont{Xu}},
  \bibinfo{author}{\bibfnamefont{Y.}~\bibnamefont{Moritomo}},
  \bibinfo{author}{\bibfnamefont{N.}~\bibnamefont{Hamada}}, \bibnamefont{and}
  \bibinfo{author}{\bibfnamefont{Y.}~\bibnamefont{Aiura}},
  \bibinfo{journal}{Phys. Rev. B} \textbf{\bibinfo{volume}{66}},
  \bibinfo{pages}{035112} (\bibinfo{year}{2002}).

\bibitem[{\citenamefont{Sarma et~al.}(2000)\citenamefont{Sarma, Mahadevan,
  Saha-Dasgupta, Ray, and Kumar}}]{sarma00}
\bibinfo{author}{\bibfnamefont{D.~D.} \bibnamefont{Sarma}},
  \bibinfo{author}{\bibfnamefont{P.}~\bibnamefont{Mahadevan}},
  \bibinfo{author}{\bibfnamefont{T.}~\bibnamefont{Saha-Dasgupta}},
  \bibinfo{author}{\bibfnamefont{S.}~\bibnamefont{Ray}}, \bibnamefont{and}
  \bibinfo{author}{\bibfnamefont{A.}~\bibnamefont{Kumar}},
  \bibinfo{journal}{Phys. Rev. Lett.} \textbf{\bibinfo{volume}{85}},
  \bibinfo{pages}{2549} (\bibinfo{year}{2000}).

\bibitem[{\citenamefont{Fang et~al.}(2001)\citenamefont{Fang, Terakura, and
  Kanamori}}]{fang01}
\bibinfo{author}{\bibfnamefont{Z.}~\bibnamefont{Fang}},
  \bibinfo{author}{\bibfnamefont{K.}~\bibnamefont{Terakura}}, \bibnamefont{and}
  \bibinfo{author}{\bibfnamefont{J.}~\bibnamefont{Kanamori}},
  \bibinfo{journal}{Phys. Rev. B} \textbf{\bibinfo{volume}{63}},
  \bibinfo{pages}{180407(R)} (\bibinfo{year}{2001}).

\bibitem[{\citenamefont{Szoteck et~al.}(2003)\citenamefont{Szoteck, Temmerman,
  Svane, Petit, and Winter}}]{szotek03}
\bibinfo{author}{\bibfnamefont{Z.}~\bibnamefont{Szoteck}},
  \bibinfo{author}{\bibfnamefont{W.}~\bibnamefont{Temmerman}},
  \bibinfo{author}{\bibfnamefont{A.}~\bibnamefont{Svane}},
  \bibinfo{author}{\bibfnamefont{L.}~\bibnamefont{Petit}}, \bibnamefont{and}
  \bibinfo{author}{\bibfnamefont{H.}~\bibnamefont{Winter}},
  \bibinfo{journal}{Phys. Rev. B} \textbf{\bibinfo{volume}{68}},
  \bibinfo{pages}{104411} (\bibinfo{year}{2003}).

\bibitem[{\citenamefont{Ray et~al.}(2001)\citenamefont{Ray, Kumar, Sarma,
  Cimino, Turchini, Zennaro, , and Zema}}]{ray01}
\bibinfo{author}{\bibfnamefont{S.}~\bibnamefont{Ray}},
  \bibinfo{author}{\bibfnamefont{A.}~\bibnamefont{Kumar}},
  \bibinfo{author}{\bibfnamefont{D.~D.} \bibnamefont{Sarma}},
  \bibinfo{author}{\bibfnamefont{R.}~\bibnamefont{Cimino}},
  \bibinfo{author}{\bibfnamefont{S.}~\bibnamefont{Turchini}},
  \bibinfo{author}{\bibfnamefont{S.}~\bibnamefont{Zennaro}}, ,
  \bibnamefont{and} \bibinfo{author}{\bibfnamefont{N.}~\bibnamefont{Zema}},
  \bibinfo{journal}{Phys. Rev. Lett.} \textbf{\bibinfo{volume}{87}},
  \bibinfo{pages}{097204} (\bibinfo{year}{2001}).

\bibitem[{\citenamefont{Balcells et~al.}(2001)\citenamefont{Balcells, Navarro,
  Bibes, Roig, Mart\'{\i}nez, and Fontcuberta}}]{balcells01}
\bibinfo{author}{\bibfnamefont{L.}~\bibnamefont{Balcells}},
  \bibinfo{author}{\bibfnamefont{J.}~\bibnamefont{Navarro}},
  \bibinfo{author}{\bibfnamefont{M.}~\bibnamefont{Bibes}},
  \bibinfo{author}{\bibfnamefont{A.}~\bibnamefont{Roig}},
  \bibinfo{author}{\bibfnamefont{B.}~\bibnamefont{Mart\'{\i}nez}},
  \bibnamefont{and}
  \bibinfo{author}{\bibfnamefont{J.}~\bibnamefont{Fontcuberta}},
  \bibinfo{journal}{Appl. Phys. Lett.} \textbf{\bibinfo{volume}{78}},
  \bibinfo{pages}{781} (\bibinfo{year}{2001}).

\bibitem[{\citenamefont{Navarro
  et~al.}(2001{\natexlab{a}})\citenamefont{Navarro, Balcells, Sandiumenge,
  Bibes, Roig, Mart\'{\i}nez, and Fontcuberta}}]{navarro01}
\bibinfo{author}{\bibfnamefont{J.}~\bibnamefont{Navarro}},
  \bibinfo{author}{\bibfnamefont{L.}~\bibnamefont{Balcells}},
  \bibinfo{author}{\bibfnamefont{F.}~\bibnamefont{Sandiumenge}},
  \bibinfo{author}{\bibfnamefont{M.}~\bibnamefont{Bibes}},
  \bibinfo{author}{\bibfnamefont{A.}~\bibnamefont{Roig}},
  \bibinfo{author}{\bibfnamefont{B.}~\bibnamefont{Mart\'{\i}nez}},
  \bibnamefont{and}
  \bibinfo{author}{\bibfnamefont{J.}~\bibnamefont{Fontcuberta}},
  \bibinfo{journal}{J. Phys.:Condens. Matter} \textbf{\bibinfo{volume}{13}},
  \bibinfo{pages}{8481} (\bibinfo{year}{2001}{\natexlab{a}}).

\bibitem[{\citenamefont{S\'anchez et~al.}(2002)\citenamefont{S\'anchez,
  Garc\'{\i}a-Hern\'andez, Mart\'{\i}nez, Alonso, Mart\'{\i}nez-Lope, Casais,
  and Mellergard}}]{sanchez02}
\bibinfo{author}{\bibfnamefont{D.}~\bibnamefont{S\'anchez}},
  \bibinfo{author}{\bibfnamefont{M.}~\bibnamefont{Garc\'{\i}a-Hern\'andez}},
  \bibinfo{author}{\bibfnamefont{J.}~\bibnamefont{Mart\'{\i}nez}},
  \bibinfo{author}{\bibfnamefont{J.}~\bibnamefont{Alonso}},
  \bibinfo{author}{\bibfnamefont{M.}~\bibnamefont{Mart\'{\i}nez-Lope}},
  \bibinfo{author}{\bibfnamefont{M.}~\bibnamefont{Casais}}, \bibnamefont{and}
  \bibinfo{author}{\bibfnamefont{A.}~\bibnamefont{Mellergard}},
  \bibinfo{journal}{J. Magn. Magnetic Mater.} \textbf{\bibinfo{volume}{242}},
  \bibinfo{pages}{729} (\bibinfo{year}{2002}).

\bibitem[{\citenamefont{Navarro et~al.}(2003)\citenamefont{Navarro, Nogu\'es,
  Mu{\~n}oz, and Fontcuberta}}]{navarro03}
\bibinfo{author}{\bibfnamefont{J.}~\bibnamefont{Navarro}},
  \bibinfo{author}{\bibfnamefont{J.}~\bibnamefont{Nogu\'es}},
  \bibinfo{author}{\bibfnamefont{J.}~\bibnamefont{Mu{\~n}oz}},
  \bibnamefont{and}
  \bibinfo{author}{\bibfnamefont{J.}~\bibnamefont{Fontcuberta}},
  \bibinfo{journal}{Phys. Rev. B} \textbf{\bibinfo{volume}{67}},
  \bibinfo{pages}{174416} (\bibinfo{year}{2003}).

\bibitem[{\citenamefont{Sher et~al.}(2005)\citenamefont{Sher, Venimadhav,
  Blamire, Kamenev, and Attfield}}]{sher05}
\bibinfo{author}{\bibfnamefont{F.}~\bibnamefont{Sher}},
  \bibinfo{author}{\bibfnamefont{A.}~\bibnamefont{Venimadhav}},
  \bibinfo{author}{\bibfnamefont{M.~G.} \bibnamefont{Blamire}},
  \bibinfo{author}{\bibfnamefont{K.}~\bibnamefont{Kamenev}}, \bibnamefont{and}
  \bibinfo{author}{\bibfnamefont{J.~P.} \bibnamefont{Attfield}},
  \bibinfo{journal}{Chemistry of Materials} \textbf{\bibinfo{volume}{17}},
  \bibinfo{pages}{176} (\bibinfo{year}{2005}).

\bibitem[{\citenamefont{Tovar et~al.}(2002)\citenamefont{Tovar, Causa, Butera,
  Navarro, Mart\'{\i}nez, Fontcuberta, and Passeggi}}]{tovar02}
\bibinfo{author}{\bibfnamefont{M.}~\bibnamefont{Tovar}},
  \bibinfo{author}{\bibfnamefont{M.}~\bibnamefont{Causa}},
  \bibinfo{author}{\bibfnamefont{A.}~\bibnamefont{Butera}},
  \bibinfo{author}{\bibfnamefont{J.}~\bibnamefont{Navarro}},
  \bibinfo{author}{\bibfnamefont{B.}~\bibnamefont{Mart\'{\i}nez}},
  \bibinfo{author}{\bibfnamefont{J.}~\bibnamefont{Fontcuberta}},
  \bibnamefont{and} \bibinfo{author}{\bibfnamefont{M.~C.~G.}
  \bibnamefont{Passeggi}}, \bibinfo{journal}{Phys. Rev. B}
  \textbf{\bibinfo{volume}{66}}, \bibinfo{pages}{024409}
  (\bibinfo{year}{2002}).

\bibitem[{\citenamefont{{Das Sarma}
  et~al.}(2003{\natexlab{a}})\citenamefont{{Das Sarma}, Hwang, and
  Kaminski}}]{dassarma03GaMnAs}
\bibinfo{author}{\bibfnamefont{S.}~\bibnamefont{{Das Sarma}}},
  \bibinfo{author}{\bibfnamefont{E.}~\bibnamefont{Hwang}}, \bibnamefont{and}
  \bibinfo{author}{\bibfnamefont{A.}~\bibnamefont{Kaminski}},
  \bibinfo{journal}{Phys. Rev. B} \textbf{\bibinfo{volume}{67}},
  \bibinfo{pages}{155201} (\bibinfo{year}{2003}{\natexlab{a}}).

\bibitem[{\citenamefont{{Das Sarma}
  et~al.}(2003{\natexlab{b}})\citenamefont{{Das Sarma}, Hwang, and
  Kaminski}}]{dassarma03GaMnAsSSC}
\bibinfo{author}{\bibfnamefont{S.}~\bibnamefont{{Das Sarma}}},
  \bibinfo{author}{\bibfnamefont{E.}~\bibnamefont{Hwang}}, \bibnamefont{and}
  \bibinfo{author}{\bibfnamefont{A.}~\bibnamefont{Kaminski}},
  \bibinfo{journal}{Solid State Communications} \textbf{\bibinfo{volume}{127}},
  \bibinfo{pages}{99} (\bibinfo{year}{2003}{\natexlab{b}}).

\bibitem[{\citenamefont{Zener}(1951)}]{zener}
\bibinfo{author}{\bibfnamefont{C.}~\bibnamefont{Zener}},
  \bibinfo{journal}{Phys. Rev.}  (\bibinfo{year}{1951}).

\bibitem[{\citenamefont{Mart\'{\i}nez et~al.}(2000)\citenamefont{Mart\'{\i}nez,
  Navarro, Balcells, and Fontcuberta}}]{martinez00}
\bibinfo{author}{\bibfnamefont{B.}~\bibnamefont{Mart\'{\i}nez}},
  \bibinfo{author}{\bibfnamefont{J.}~\bibnamefont{Navarro}},
  \bibinfo{author}{\bibfnamefont{L.}~\bibnamefont{Balcells}}, \bibnamefont{and}
  \bibinfo{author}{\bibfnamefont{J.}~\bibnamefont{Fontcuberta}},
  \bibinfo{journal}{J. Phys.:Condens. Matter} \textbf{\bibinfo{volume}{12}},
  \bibinfo{pages}{10515} (\bibinfo{year}{2000}).

\bibitem[{\citenamefont{Navarro
  et~al.}(2001{\natexlab{b}})\citenamefont{Navarro, Frontera, Balcells,
  Mart\'{\i}nez, and Fontcuberta}}]{navarro01PRB}
\bibinfo{author}{\bibfnamefont{J.}~\bibnamefont{Navarro}},
  \bibinfo{author}{\bibfnamefont{C.}~\bibnamefont{Frontera}},
  \bibinfo{author}{\bibfnamefont{L.}~\bibnamefont{Balcells}},
  \bibinfo{author}{\bibfnamefont{B.}~\bibnamefont{Mart\'{\i}nez}},
  \bibnamefont{and}
  \bibinfo{author}{\bibfnamefont{J.}~\bibnamefont{Fontcuberta}},
  \bibinfo{journal}{Phys. Rev. B} \textbf{\bibinfo{volume}{64}},
  \bibinfo{pages}{092411} (\bibinfo{year}{2001}{\natexlab{b}}).

\bibitem[{\citenamefont{Frontera et~al.}(2003)\citenamefont{Frontera, Rubi,
  Navarro, no, Fontcuberta, and Ritter}}]{frontera03}
\bibinfo{author}{\bibfnamefont{C.}~\bibnamefont{Frontera}},
  \bibinfo{author}{\bibfnamefont{D.}~\bibnamefont{Rubi}},
  \bibinfo{author}{\bibfnamefont{J.}~\bibnamefont{Navarro}},
  \bibinfo{author}{\bibfnamefont{J.~L. G.-M.} \bibnamefont{no}},
  \bibinfo{author}{\bibfnamefont{J.}~\bibnamefont{Fontcuberta}},
  \bibnamefont{and} \bibinfo{author}{\bibfnamefont{C.}~\bibnamefont{Ritter}},
  \bibinfo{journal}{Phys. Rev. B} \textbf{\bibinfo{volume}{68}},
  \bibinfo{pages}{012412} (\bibinfo{year}{2003}).

\bibitem[{\citenamefont{Fontcuberta et~al.}(2005)\citenamefont{Fontcuberta,
  Rubi, Frontera, noz, Wojcik, amd S.~Nadolski, Izquierdo, Avila, and
  Asensio}}]{fontcuberta05}
\bibinfo{author}{\bibfnamefont{J.}~\bibnamefont{Fontcuberta}},
  \bibinfo{author}{\bibfnamefont{D.}~\bibnamefont{Rubi}},
  \bibinfo{author}{\bibfnamefont{C.}~\bibnamefont{Frontera}},
  \bibinfo{author}{\bibfnamefont{J.~L. G.-M.} \bibnamefont{noz}},
  \bibinfo{author}{\bibfnamefont{M.}~\bibnamefont{Wojcik}},
  \bibinfo{author}{\bibfnamefont{E.~J.} \bibnamefont{amd S.~Nadolski}},
  \bibinfo{author}{\bibfnamefont{M.}~\bibnamefont{Izquierdo}},
  \bibinfo{author}{\bibfnamefont{J.}~\bibnamefont{Avila}}, \bibnamefont{and}
  \bibinfo{author}{\bibfnamefont{M.}~\bibnamefont{Asensio}},
  \bibinfo{journal}{J. Magn. Magnetic Mater.}
  \textbf{\bibinfo{volume}{290-291}}, \bibinfo{pages}{974}
  (\bibinfo{year}{2005}).

\bibitem[{\citenamefont{Navarro et~al.}(2004)\citenamefont{Navarro,
  Fontcuberta, Izquierdo, Avila, and Asensio}}]{navarro04}
\bibinfo{author}{\bibfnamefont{J.}~\bibnamefont{Navarro}},
  \bibinfo{author}{\bibfnamefont{J.}~\bibnamefont{Fontcuberta}},
  \bibinfo{author}{\bibfnamefont{M.}~\bibnamefont{Izquierdo}},
  \bibinfo{author}{\bibfnamefont{J.}~\bibnamefont{Avila}}, \bibnamefont{and}
  \bibinfo{author}{\bibfnamefont{M.}~\bibnamefont{Asensio}},
  \bibinfo{journal}{Phys. Rev. B} \textbf{\bibinfo{volume}{69}},
  \bibinfo{pages}{115101} (\bibinfo{year}{2004}).

\bibitem[{\citenamefont{Phillips et~al.}(2003)\citenamefont{Phillips,
  Chattopadhyay, and Millis}}]{millis03}
\bibinfo{author}{\bibfnamefont{K.}~\bibnamefont{Phillips}},
  \bibinfo{author}{\bibfnamefont{A.}~\bibnamefont{Chattopadhyay}},
  \bibnamefont{and} \bibinfo{author}{\bibfnamefont{A.~J.}
  \bibnamefont{Millis}}, \bibinfo{journal}{Phys. Rev. B}
  \textbf{\bibinfo{volume}{67}}, \bibinfo{pages}{125119}
  (\bibinfo{year}{2003}).

\bibitem[{\citenamefont{Alonso et~al.}(2003)\citenamefont{Alonso, Fern\'andez,
  Guinea, Lesmes, and Martin-Mayor}}]{guinea03}
\bibinfo{author}{\bibfnamefont{J.}~\bibnamefont{Alonso}},
  \bibinfo{author}{\bibfnamefont{L.}~\bibnamefont{Fern\'andez}},
  \bibinfo{author}{\bibfnamefont{F.}~\bibnamefont{Guinea}},
  \bibinfo{author}{\bibfnamefont{F.}~\bibnamefont{Lesmes}}, \bibnamefont{and}
  \bibinfo{author}{\bibfnamefont{V.}~\bibnamefont{Martin-Mayor}},
  \bibinfo{journal}{Phys. Rev. B} \textbf{\bibinfo{volume}{67}},
  \bibinfo{pages}{214423} (\bibinfo{year}{2003}).

\bibitem[{\citenamefont{Taraphder and Guinea}(2004)}]{TG04}
\bibinfo{author}{\bibfnamefont{A.}~\bibnamefont{Taraphder}} \bibnamefont{and}
  \bibinfo{author}{\bibfnamefont{F.}~\bibnamefont{Guinea}},
  \bibinfo{journal}{Phys. Rev. B} \textbf{\bibinfo{volume}{70}},
  \bibinfo{pages}{224438} (\bibinfo{year}{2004}).

\bibitem[{\citenamefont{de~Gennes}(1960)}]{degennes}
\bibinfo{author}{\bibfnamefont{P.-G.} \bibnamefont{de~Gennes}},
  \bibinfo{journal}{Phys. Rev.} \textbf{\bibinfo{volume}{118}},
  \bibinfo{pages}{141} (\bibinfo{year}{1960}).

\bibitem[{DMF()}]{DMFT}
\bibinfo{note}{Note that this scheme is somewhat different from the one used in
  Eq.~(\ref{DMFT_eq}), which is based on the expectation value of $\langle \cos
  ( \theta / 2 ) \rangle$ estimated from Dynamical Mean Field Theory. Both
  schemes lead to very similar results, as $2/3 \approx 1/\sqrt{2}$.}

\bibitem[{\citenamefont{Guinea et~al.}(2000)\citenamefont{Guinea,
  G\'omez-Santos, and Arovas}}]{GGA00}
\bibinfo{author}{\bibfnamefont{F.}~\bibnamefont{Guinea}},
  \bibinfo{author}{\bibfnamefont{G.}~\bibnamefont{G\'omez-Santos}},
  \bibnamefont{and} \bibinfo{author}{\bibfnamefont{D.}~\bibnamefont{Arovas}},
  \bibinfo{journal}{Phys. Rev. B} \textbf{\bibinfo{volume}{62}},
  \bibinfo{pages}{391} (\bibinfo{year}{2000}).

\bibitem[{\citenamefont{Chattopadhyay
  et~al.}(2001{\natexlab{a}})\citenamefont{Chattopadhyay, {Das Sarma}, and
  Millis}}]{chattopadhyay01GaMnAs}
\bibinfo{author}{\bibfnamefont{A.}~\bibnamefont{Chattopadhyay}},
  \bibinfo{author}{\bibfnamefont{S.}~\bibnamefont{{Das Sarma}}},
  \bibnamefont{and} \bibinfo{author}{\bibfnamefont{A.~J.}
  \bibnamefont{Millis}}, \bibinfo{journal}{Phys. Rev. Lett.}
  \textbf{\bibinfo{volume}{87}}, \bibinfo{pages}{227202}
  (\bibinfo{year}{2001}{\natexlab{a}}).

\bibitem[{\citenamefont{Chattopadhyay
  et~al.}(2001{\natexlab{b}})\citenamefont{Chattopadhyay, {Das Sarma}, and
  Millis}}]{chattopadhyay01PRB}
\bibinfo{author}{\bibfnamefont{A.}~\bibnamefont{Chattopadhyay}},
  \bibinfo{author}{\bibfnamefont{S.}~\bibnamefont{{Das Sarma}}},
  \bibnamefont{and} \bibinfo{author}{\bibfnamefont{A.~J.}
  \bibnamefont{Millis}}, \bibinfo{journal}{Phys. Rev. B}
  \textbf{\bibinfo{volume}{64}}, \bibinfo{pages}{012416}
  (\bibinfo{year}{2001}{\natexlab{b}}).

\bibitem[{\citenamefont{Calder\'on and Brey}(1998)}]{calderonMC98}
\bibinfo{author}{\bibfnamefont{M.~J.} \bibnamefont{Calder\'on}}
  \bibnamefont{and} \bibinfo{author}{\bibfnamefont{L.}~\bibnamefont{Brey}},
  \bibinfo{journal}{Phys. Rev. B} \textbf{\bibinfo{volume}{58}},
  \bibinfo{pages}{3286} (\bibinfo{year}{1998}).

\end{thebibliography}

\end{document}